\documentclass[aps,prd,notitlepage,showpacs,nofootinbib,superscriptaddress]{revtex4-1}

\usepackage{graphicx}
\usepackage[utf8]{inputenc} 

\usepackage{amsmath}
\usepackage{amssymb}
\usepackage{float}
\usepackage{comment}
\usepackage{slashed}
\usepackage[normalem]{ulem}

\usepackage{booktabs}
\usepackage{mathtools,braket,blkarray}

\usepackage[usenames,dvipsnames]{color}
\usepackage[colorinlistoftodos]{todonotes}
\usepackage[colorlinks=true,citecolor=darkred,urlcolor=darkred, pdfborder={0 0 0}]{hyperref}
\usepackage[normalem]{ulem}

\usepackage{multirow}
\definecolor{darkred}{rgb}{0.6,0,0}
\usepackage[colorinlistoftodos]{todonotes}

\definecolor{linkcolor}{rgb}{0,0,0.5}
 
 \newcommand {\ignore}[1]{}

\newcommand{\mg}[1]{{\color{purple}[\underline{\bf MG}:~{\bf #1}]}}

\newcommand{\FC}[1]{{\color{green}#1}}

\def\gsim{\raise0.3ex\hbox{$\;>$\kern-0.75em\raise-1.1ex\hbox{$\sim\;$}}}
\def\lsim{\raise0.3ex\hbox{$\;<$\kern-0.75em\raise-1.1ex\hbox{$\sim\;$}}}

\usepackage{soul}

\definecolor{mightnightblue}{RGB}{25,25,112}

\definecolor{brown}{rgb}{0.59, 0.29, 0.0}

\def\21{$\mathrm{SU(2)_L \otimes U(1)_Y}$}

\begin{document}

\title{\boldmath \color{BrickRed} Probing high-energy solar axion flux with a large scintillation neutrino detector
}

\author{Giuseppe Lucente}
\email{giuseppe.lucente@ba.infn.it}
\affiliation{Dipartimento Interateneo di Fisica “Michelangelo Merlin”, Via Amendola 173, 70126 Bari, Italy}
\affiliation{Istituto Nazionale di Fisica Nucleare - Sezione di Bari, Via Orabona 4, 70126 Bari, Italy}%

\author{Newton Nath}\email{newton.nath@ba.infn.it}
\affiliation{Istituto Nazionale di Fisica Nucleare - Sezione di Bari, Via Orabona 4, 70126 Bari, Italy}

\author{Francesco Capozzi}
\email{francesco.capozzi@univaq.it}
\affiliation{Dipartimento di Scienze Fisiche e Chimiche, Universit\`a degli Studi dell'Aquila, 67100 L’Aquila, Italy}
  
\author{Maurizio Giannotti}\email{mgiannotti@barry.edu}
\affiliation{Department of Chemistry and Physics, Barry University, 11300 NE 2nd Ave., Miami Shores, FL 33161, USA}%

\author{Alessandro Mirizzi}
\email{alessandro.mirizzi@ba.infn.it}
\affiliation{Dipartimento Interateneo di Fisica “Michelangelo Merlin”, Via Amendola 173, 70126 Bari, Italy}
\affiliation{Istituto Nazionale di Fisica Nucleare - Sezione di Bari, Via Orabona 4, 70126 Bari, Italy}%


\begin{abstract}
\vspace{0.5cm}
\noindent
We investigate the 5.49 MeV solar axions flux produced in the $p(d,$ $\!\! ^3{\rm He})a$ 
reaction and analyze the potential to detect it with the forthcoming large underground neutrino oscillation experiment Jiangmen Underground Neutrino Observatory (JUNO). 
The JUNO detector could reveal axions through various processes such as Compton and inverse Primakoff conversion, as well as through their decay into two photons or  electron-positron pairs inside the detector.
We perform a detailed numerical analysis in order to forecast the sensitivity  on different combinations of the axion-electron ($ g_{ae} $), axion-photon ($g_{a\gamma} $), and
isovector axion-nucleon ($ g_{3aN} $) 
couplings, using the expected JUNO data for different benchmark values of axion mass in a model-independent way. We find that JUNO would improve by approximately one order of magnitude current bounds by Borexino and it has the best sensitivity among neutrino experiments.
\end{abstract}
\maketitle

\section{Introduction}

The theory of the strong interactions, quantum chromodynamics (QCD) is expected to violate the charge-conjugation parity (CP) symmetry. 
However, all experimental observations are compatible with CP conservation in the strong interactions. 
Explaining the observed smallness of the CP violation in QCD remains, after several decades, an unresolved puzzle in particle physics, known as the \emph{strong CP problem}. 
The most cogent solution of this problem is to postulate 
an anomalous global U(1) symmetry -- the Peccei-Quinn (PQ) symmetry -- that is broken spontaneously, leading to a pseudo-Nambu-Goldstone boson (pNGB) called the QCD axion~\cite{Peccei:1977hh,Wilczek:1977pj,Weinberg:1977ma}.
At the current juncture, the axion is one of the best motivated elementary particles beyond the Standard Model (SM).
In fact, in addition to providing the most appealing explanation for the strong CP problem, axions are also excellent dark matter candidates~\cite{Dine:1982ah,Abbott:1982af,Preskill:1982cy,Davis:1986xc,Kim:2008hd,Marsh:2015xka,Adams:2022pbo,Berti:2022rwn}.
%
The theory allows for many different realizations of  QCD axion models, with very specific phenomenology (see Ref.~\cite{DiLuzio:2020wdo} for a comprehensive review). 

The model-dependent axion couplings to SM fields open up strategies for their detection.
The most accessible experimental detection channels are through the couplings with photons ($ g_{a\gamma} $),  electrons ($ g_{a e} $), and  nucleons (isosinglet $ g_{0aN} $ and isotriplet  $ g_{3aN} $).
These interactions are represented in the effective low-energy axion Lagrangian

\begin{align}
  \label{eq:AxionInter}
  \mathcal{L} = \dfrac{1}{2} \left( \partial_{\mu} a \right)^2  - m^2_a a^2 
  - \frac{1}{4}\,g_{a\gamma}\, a F_{\mu\nu}\widetilde{F}^{\mu\nu}
  - ig_{ae}a\,\bar e \gamma_5 e
  - ia\bar N\gamma_5\,\left(g_{0aN} + \tau_3 g_{3aN}\right)N\ ,
\end{align}
where the first two terms represent the kinetic and mass terms of the axion field $a$, 
$F_{\mu\nu}$ and $ \widetilde{F}^{\mu\nu} $ are the electromagnetic field strength tensor and its dual, {and} $N$ refers to the proton-neutron isospin doublet. 

Typical axion models are constrained to very small masses, below $\sim1$~eV. 
However, there exist non-minimal models which predict heavy axions, with masses larger than $\sim 100$~keV, without spoiling
the solution of the strong CP problem (a list of references can be found in Sec.~6.7 of Ref.~\cite{DiLuzio:2020wdo}). 
Heavy QCD axions are well motivated since they can provide a simple solution~\cite{Agrawal:2017ksf} to the axion quality problem~\cite{Kamionkowski:1992mf,Holman:1992us,Barr:1992qq,Rubakov:1997vp,Berezhiani:2000gh,Hook:2019qoh}, i.e. the explicit breaking of the $U(1)$ Peccei-Quinn symmetry by higher dimensional Planck-suppressed operators induced by quantum gravity, which could spoil the PQ mechanism. 
Besides QCD axions, (heavy) axion-like particles (ALPs) emerge in compactification scenarios of string theory \cite{Svrcek:2006yi,Arvanitaki:2009fg,Cicoli:2012sz} as well as in ``relaxion'' models~\cite{Graham:2015cka}. 
In this work, we
use the term ``axions" to refer to both QCD axions and ALPs.

A remarkable experimental effort has been devoted to axion searches in recent years (see Refs.~\cite{Irastorza:2018dyq,DiVecchia:2019ejf,DiLuzio:2020wdo,Agrawal:2021dbo,Sikivie:2020zpn,Giannotti:2022euq} for recent reviews and updates).
Currently, experimental searches have started the exploration of large sections of the parameter space allowed by astrophysical considerations~\cite{Giannotti:2017hny,Agrawal:2021dbo,DiLuzio:2021ysg}, generating excitement and hopes for discovery in the next decade or so~\cite{Giannotti:2017law,Giannotti:2022euq}.\footnote{Updated plots on axion experimental limits, as well as on other phenomenological bounds, can be found in Ref.~\cite{AxionLimits}.}
Here, our focus will be on studying solar axions. 
The Sun is one of the most important natural sources of axions. 
In the hot core, $T_c\sim 1\,$ keV, axions of mass below a few keV and with thermal energies can be efficiently  produced through processes involving the coupling to photons $g_{a\gamma}$, i.e., the Primakoff effect~\cite{Raffelt:1985nk,Raffelt:1987np} and photon-axion conversions in the solar magnetic field~\cite{Caputo:2020quz,OHare:2020wum,Guarini:2020hps},
or the coupling to electrons $g_{ae}$,  such as electrons scattering off nuclei,
electron bremsstrahlung and Compton effect~\cite{Redondo:2013wwa} 
(see Ref.~\cite{Hoof:2021mld} for details and updated rates). 
A stringent constraint on solar axions coupled to photons
was placed a few years ago by the CERN Axion Solar Experiment (CAST)~\cite{CAST:2017uph},
which excluded the couplings
 $ g_{a\gamma} > 0.66 \times 10^{-10}$ GeV$ ^{-1} $ at 95\% confidence level for $m_a\lesssim 0.02$~eV.
A similar bound can be derived from observations of horizontal branch (HB) stars in globular clusters~\cite{Ayala:2014pea,Straniero:2015nvc}. 
 A more recent analysis~\cite{Dolan:2022kul} 
found a slightly stronger bound, specifically $g_{a\gamma}\lesssim 0.34 \times 10^{-10}$~GeV$^{-1}$ for $m_a<1$~keV, by measuring the ratio of stars in the asymptotic giant branch and in the HB in globular clusters.

The very low mass region ($m_a\lesssim {\rm neV}$) is subject to considerably more severe constraints from various astrophysical observations (see, e.g., Refs.\cite{Wouters:2013hua,Marsh:2017yvc,Reynolds:2019uqt,Dessert:2020lil,Xiao:2020pra,Reynes:2021bpe,Calore:2021hhn,Dessert:2021bkv,Xiao:2022rxk}) and is a target for several proposed laboratory searches, e.g., ABRACADABRA~\cite{Kahn:2016aff}. 
On the other hand, at much larger masses ($m_a\sim$ keV), X-ray observations by NuSTAR~\cite{NuSTAR:2013yza} have been used to constrain axions trapped in the gravitational potential of the Sun, forming the ``Solar basin'', leading to very strong  bounds on $g_{a\gamma}$ and $g_{ae}$ for $m_a\sim O(10)$~keV~\cite{DeRocco:2022jyq}.
In addition, the flux generated from the axion-electron coupling can be searched by a new generation of axion helioscope experiments, BabyIAXO and IAXO~\cite{IAXO:2019mpb,IAXO:2020wwp}, which are sensitive to the product of couplings $g_{ae}\times g_{a \gamma}$, and by underground dark matter experiments such as Xenon~\cite{XENON:2020rca}, LUX~\cite{LUX:2017glr}, and PandaX-II~\citep{Fu:2017lfc}. 
However, the current experimental bounds are not competitive with other astrophysical constraints, in particular with the bounds form red giant stars~\cite{Straniero:2020iyi,Capozzi:2020cbu}.
Finally, the axion couplings with nuclei, $g_{0aN}$ and $g_{3aN}$,
also  contribute to the solar axion flux, through nuclear reactions with an axion in the final state, or through de-excitation of nuclei. 
These are non-thermal processes which   
produce an almost monocromatic axion spectrum. Since the mass limitation of a few keV
 for thermal axions does not apply to these processes, they can probe higher masses.

The study of axions from nuclear reactions has a long history, as it was originally considered one of the most efficient way to hunt for these particles~\cite{Donnelly:1978ty}. 
An early attempt to study the axion flux from nuclear processes in the Sun can be found in Ref.~\cite{Raffelt:1982dr}, and searches for resonant absorption of solar axions emitted in the nuclear magnetic transitions have been performed with $^{57}$Fe~\cite{Moriyama:1995bz,Krcmar:1998xn,CAST:2009jdc,CUORE:2012ymr}, $^7$Li \cite{Derbin:2005xc,Belli:2008zzb,Borexino:2008wiu,CAST:2009klq} and $^{83}$Kr nuclei~\cite{Gavrilyuk:2014mch,Jakovcic:2004sh}.
%
Recent studies include helioscope sensitivity to different nuclear processes~\cite{CAST:2009klq,CAST:2009jdc,DiLuzio:2021qct} and a bound on $g_{3aN}$ from SNO data, $ g_{3aN} \gtrsim 2 \times 10^{-5}$ (95\% confidence level)~\cite{Bhusal:2020bvx}. Axions from the $p + d \rightarrow\rm{^3He}+ a(5.49 \,\  MeV)$ reaction have been probed by Borexino, constraining the coupling combinations $(g_{3aN},~g_{ae})$ and $(g_{3aN},~g_{a\gamma})$~\cite{Borexino:2012guz}. In this work, our goal is to improve the limits set by Borexino using other neutrino experiments. In principle, with respect to \cite{Borexino:2012guz}, Borexino has now collected its full dataset \cite{Borexino:2017uhp}, with which most likely the bounds obtained in \cite{Borexino:2012guz} can be improved. However, an analysis of such dataset is challenging when performed outside of collaboration, especially considering that the detector has been changing significantly over time and the information of this time evolution is not entirely available. Another detector sensitive to MeV neutrinos is Super-Kamiokande~\cite{Super-Kamiokande:2016yck}, whose 22 kton of fiducial mass represents a factor $\sim 200$ improvement with respect to the 100 tons of Borexino. Nevertheless, since we are dealing with the search of monochromatic axions from the Sun, the energy resolution is a key factor, and the better one achieved in Borexino   compensates for the relatively low fiducial volume. In terms of future detectors, the proposed Jiangmen Underground Neutrino Observatory (JUNO) represents a step-forward, since it combines a large fiducial mass ($\sim20$ kton) and an exquisite energy resolution (3\%/$\sqrt{E({\rm MeV}}$)~\cite{JUNO:2015zny}. For this reason, in this work, we focus on the JUNO detector, showing that it can improve by about an order of magnitude bounds set by the Borexino.
In order to serve our purpose, we consider the JUNO events spectrum provided in Ref.~\cite{JUNO:2020hqc}.
In particular, we focus on the axion flux from the $p + d \rightarrow\rm{^3He}+ a(5.49 \,\ MeV)$ reaction, as in Ref.~\cite{Borexino:2012guz,Bhusal:2020bvx}.
This  flux can be detected through various processes.  The most relevant are the Compton conversion of axions to photons, ${\rm a}+e\rightarrow e+\gamma$, inverse Primakoff conversion on nuclei, ${\rm a}+Z\rightarrow\gamma+Z$, axion electron-pair production, and axions decay into two photons as well as two electrons.

We organise the manuscript as follows. In Sec.~\ref{sec:influx} we describe the solar axion flux at the source and at the Earth. In Sec.~\ref{sec:JUNOdet} we discuss how to evaluate the solar axion event rates in JUNO, giving details on the experimental set-up and the possible axion interactions in the detector. In Sec.~\ref{sec:sensitivity} we compute JUNO sensitivity, while in Sec.~\ref{sec:HK} we compare JUNO with other current and future neutrino experiments. Finally, conclusions are summarized in Sec.~\ref{sec:Conclusion}. A  discussion of the supernova bound on the axion-nucleon coupling is presented in the Appendix.




\section{High-energy solar axion flux}
\label{sec:influx}
As discussed in the previous section, axions can be non-thermally produced in the Sun through nuclear reaction processes  
induced by the last term in Eq.~\eqref{eq:AxionInter} (see, e.g., Refs.~\cite{Vergados:2021ejk,Massarczyk:2021dje} for updated studies). 
A monochromatic flux of axions is expected to be produced in magnetic dipole transitions from the de-excitation of excited levels of nuclei in the Sun, 
e.g. $^{57}{\rm Fe}^* \to\, ^{57}{\rm Fe}+a({\rm 14.4\,keV})$ 
and $^{83}{\rm Kr}^* \to\, ^{83}{\rm Kr}+a({\rm 9.4\,keV})$, 
or from nuclear reactions such as $p+d\to \, ^3 {\rm He}+a\,(5.49\,{\rm MeV})$.
Detailed studies (see, e.g. Ref.~\cite{Massarczyk:2021dje} and in particular Fig.~7 therein) show that 
this last process is
one of the most efficient axion production mechanism mediated by the axion-nucleon coupling. 
This is the \emph{axionic counterpart} of the famous 
$p+d\to \, ^3 {\rm He}+\gamma$ process, responsible for the transformation of nearly all deuterium into  $^3{\rm He}$ nuclei in the Sun.
According to the Standard Solar Model,
this is the second stage of the $pp$-solar fusion chain, following the first step in which $99.7\%$ of deuterium is produced after the fusion of two protons, $p\,+\,p\,\rightarrow \,d\,+\,e^+\,+\nu_e$, 
and the remaining $0.3\%$ via the $p\,+\,p\,+e^-\rightarrow\,d\,+\nu_e$ process. 
Practically, every single deuterium produced in such a way ends up capturing a proton,
undergoing the reaction $p+d\to \, ^3 {\rm He}+\gamma$ on a time scale of O(1$\,$s). 
Though this is not the only deuterium reaction allowed in the Sun, the enormous relative abundance of protons with respect to deuterium makes reactions such as  $d+d\to p+t$
or $d+d\to n+^3$He extremely unlikely.
%
Consequently, for all practical purposes the standard model predicts one neutrino and one photon for each deuterium nucleus produced in the first stage of the $pp$ chain,
$\Phi_{\gamma pp}=\Phi_{\nu pp}$. 

If axions exist (and are coupled to nucleons), however, the second stage of the $pp$ chain may produce an axion rather than a photon. 
The number of axions produced can be related to the number of photons and thus (assuming the axions are not reabsorbed in the solar medium) to the neutrino flux.  
Specifically, 
$\Phi_{a 0}=(\Gamma_a/\Gamma_\gamma)\Phi_{\nu pp}$, where the coefficient
%
%
\begin{align}
    \frac{\Gamma_a}{\Gamma_\gamma}=
\left( \frac{k_a}{k_\gamma} \right)^3
\frac{1}{2\pi\alpha}\frac{1}{1+\delta^2}
\left[ \frac{\beta \, g_{0aN} + g_{3aN}} 
{\left( \mu_0-\frac12 \right) \beta + \mu_3 -\eta} \right]^2\, \,\ ,
\end{align}
measures the probability for a given nuclear transition to result in an axion rather
than a photon emission~\cite{Avignone:1988bv}.
Here,
$k_a$ and $k_\gamma$ are the axion and photon momenta, $\alpha$ is the electromagnetic fine structure constant, $\mu_0=\mu_p+\mu_n\approx 0.88$ and $\mu_3=\mu_p-\mu_n\approx 4.77$ are the isoscalar and isovector nuclear magnetic moments (expressed in nuclear magnetons), $\delta$ is the E2/M1 mixing ratio for the nuclear transition (E2 and M1 indicate respectively  the electric quadrupole and the magnetic dipole transition), while $\beta$ and $\eta$ are constants dependent on the nuclear structure. 
An efficient nuclear transition should have $\delta \ll 1$, so that the M1 transition dominates.
The $^3$He formation process is characterized by $\beta=\delta=0$, and $\eta=1$ (see, e.g., Table II of Ref.~\cite{Massarczyk:2021dje}) which, in particular, implies that only the isotriplet axion-nucleon coupling $ g_{3aN} $ is relevant in this process,
numerically,
\begin{equation}
 \frac{\Gamma_a}{\Gamma_\gamma}\simeq 1.53 \,g_{3aN}^2\left(\frac{k_a}{k_\gamma}\right)^3   \;.
\end{equation}

If axions interact sufficiently weakly, they will escape the Sun without being reabsorbed, just like the neutrinos, and produce an axion flux on Earth.
In this case, inserting the known $pp$ solar neutrino
flux, $\Phi_{\nu pp}=6.0\times 10^{10}\,{\rm cm}^{-2} {\rm s}^{-1}$~\cite{Serenelli:2011py,Borexino:2017rsf}, and 
accounting for a possible axion decay, 
we find  the expected axion flux on Earth
\begin{align}
    \Phi_{a} = \Phi_{a0} \,e^{-d_{\odot}/l_{\rm tot}}
    \simeq 3.23\times 10^{10}     \,e^{-d_{\odot}/l_{\rm tot}}
\,g_{3aN}^2(k_a/k_\gamma)^3
    \,{\rm cm}^{-2} {\rm s}^{-1}\,,
    \label{eq:phia}
\end{align}
where $l_{\rm tot}=(1/l_\gamma +1/l_e)^{-1}$ is the total axion decay length,
with $l_\gamma$ and $l_e$ the decay length in photons and electron pairs respectively,
and $d_{\odot}=1.5\times10^{13}$ the Earth-Sun distance. 
If the axion interactions are large enough, they can be reabsorbed in the Sun and Eq.~\eqref{eq:phia} becomes invalid. 
The axion-nucleon coupling can induce axion absorption after the axiodissociation of nuclei $a\,+\,Z \rightarrow Z_1\,+\,Z_2$. 
Axions with energy 5.49 MeV can dissociate $^{17}$O, $^{13}$C and $^2$H. 
It is possible to show that couplings  $g_{3aN}\lesssim 10^{-3}$ are required for axions not to be trapped inside the Sun~\cite{Raffelt:1982dr}. 
In addition, as discussed in Ref.~\cite{Borexino:2012guz}, axions would be  trapped in the Sun for $g_{ae}\gtrsim 10^{-6}$ or $g_{a\gamma}\gtrsim 10^{-4}$~GeV$^{-1}$ through inverse Compton and inverse Primakoff absorption respectively (Cf. Sec.~\ref{sec:Axion_interactions_and_decays}).  

The axion flux from $p+d\to \, ^3 {\rm He}+a$ 
has been explored by the Borexino~\cite{Borexino:2012guz} and by the  CAST~\cite{CAST:2009jdc,CAST:2009klq,CAST:2017uph} experiments via different detection channels related to axion couplings to photons and electrons.
Furthermore, using the deuterium ``axiodissociation'' process $a\,+\,d\rightarrow n\,+\,p$, Ref.~\cite{Bhusal:2020bvx} derived
a bound on $g_{3aN}$ through the analysis of Sudbury Neutrino Observatory (SNO) data, excluding the region $2\times 10^{-5}<g_{3aN}<10^{-3}$~\cite{Bhusal:2020bvx} for axion masses up to 5.49 MeV. 
Here, in analogy with Ref.~\cite{Borexino:2012guz}, we constrain $5.49$~MeV axions detectable in JUNO after interactions with photons and electrons.
As we shall see, the axion flux may be large enough to allow the exploration of a region of the parameter space  not yet probed by other experiments.

\section{JUNO as a detector for solar axions}
\label{sec:JUNOdet}
In this section, we describe two possibilities of our estimate of the solar axions event rates in JUNO.
First, we consider, if axions are detected after interacting with the detector, the expected number of events per unit time is given by
\begin{equation}
    N_{\rm ev} = N_{T} \otimes \Phi_a \otimes \sigma \otimes \mathcal{R} \otimes \varepsilon\,,
\label{eq:nev}    
\end{equation}
where $N_T$ is the number of targets and the initial axion flux $\Phi_a$ is convoluted with the cross section $\sigma$ in the detector, the detector energy resolution $\mathcal{R}$ and the detector efficiency $\varepsilon$.
On the other hand, if axions decay into photons or electron-positron pairs inside the detector
the event rate is evaluated as~\footnote{ Notice that in this work we consider axions interacting with nucleons and either photons or electrons, but not both. For this reason we are not combining the decay lengths of axions into photons and electrons in Eq.~\eqref{eq:nevd}. }
\begin{equation}
    N_{\rm ev} = \Phi_a \frac{V}{l_i} \varepsilon\,,
    \label{eq:nevd}
\end{equation}
where $V$ is the detector fiducial volume and $l_i$ is the decay length in the $i-$th decay channel.
In the following, we assume for all detection channels $\varepsilon = 1$ over the energy threshold.

\subsection{Experimental set-up}

JUNO is a multi-purpose underground liquid scintillator (LS) detector, whose primary physics goal is to determine the neutrino mass ordering
(see, e.g., \cite{JUNO:2015zny} for a review on the detector physics case), thanks to its excellent energy resolution capability and the large fiducial volume.\footnote{The JUNO detector can be used to test various new physics predictions, such as proton decay, neutrino non-standard interactions, and violation of Lorentz invariance.}
The main features of the JUNO detector have been thoroughly described in Ref.~\cite{JUNO:2015zny}. 
It consists of a central detector, a water-Cherenkov detector, and a muon tracker.
The central detector is a Liquid Scintillator (LS) of 20 kton fiducial mass with 
energy resolution 
\begin{equation}
    \frac{\sigma}{E} = \frac{3\%}{\sqrt{E}}\,,
\label{eq:eres}
\end{equation} 
where both $\sigma$ and $E$ are expressed in MeV. The detector is made of Linear
alkylbenzene (LAB), C$_{19}$H$_{32}$, doped with $3$ g/L of $2,5$-diphenyloxazole (PPO) and $15$ mg/L of p-bis-(o-methylstyryl)-benzene (bis-MSB). The density of the LS is  $0.859$ g/ml and it is contained in a spherical container of radius $17.7$ m, sorrounded by $\sim 53000$ multipliers~\cite{JUNO:2015zny,JUNO:2022lpc}.

As discussed in Ref.~\cite{JUNO:2020hqc}, in order to reduce the background and detect $^8$B solar neutrinos with a threshold energy of $2$ MeV, an energy dependent fiducial volume (FV) cut is considered
\begin{itemize}
    \item FV of 7.9 kton and $r=13$~m for $2~{\rm MeV}<E\leq3~{\rm MeV}$\,,
    \item FV of 12.2 kton and $r=15$~m for $3~{\rm MeV}<E\leq5~{\rm MeV}$\,,
    \item FV of 16.2 kton and $r=16.5$~m for $E >5~{\rm MeV}$\,.
\end{itemize}
Further exclusion cuts to reduce the background are discussed in Ref.~\cite{JUNO:2020hqc}. With this experimental set-up and after applying all the cuts about 60,000 solar neutrino events and 30,000 radioactive background events are expected in 10 years of data taking (see Table 4 and Fig. 11 in Ref.~\cite{JUNO:2020hqc}). For our work, both types of events contribute to the background.
On the other hand, the axion signal induced by the coupling with electrons $g_{ae}$ and photons $g_{a\gamma}$ is produced via the processes discussed in the next Section.

\subsection{Axion detection channels}
\label{sec:Axion_interactions_and_decays}

\subsubsection{Axion-electron coupling}
Axions interacting with electrons can be detected through Compton-like scattering $a+e^-\rightarrow \gamma + e^-$~\cite{Mikaelian:1978jg,Brodsky:1986mi,Chanda:1987ax}, the axio-electric effect $a+e^-+Ze\rightarrow e^- + Ze$~\cite{Dimopoulos:1986mi,Pospelov:2008jk,Derevianko:2010kz}, pair  production in the electric field of nuclei and electrons $a+Ze\to Ze+e^-+ e^+$~\cite{Kim:1982xb,Kim:1984ss,Blumlein:1991xh}, and the decay into electron-positron pairs $a\to  e^+ + e^-$.

\begin{figure}[t]
	\includegraphics[width=0.6\textwidth]{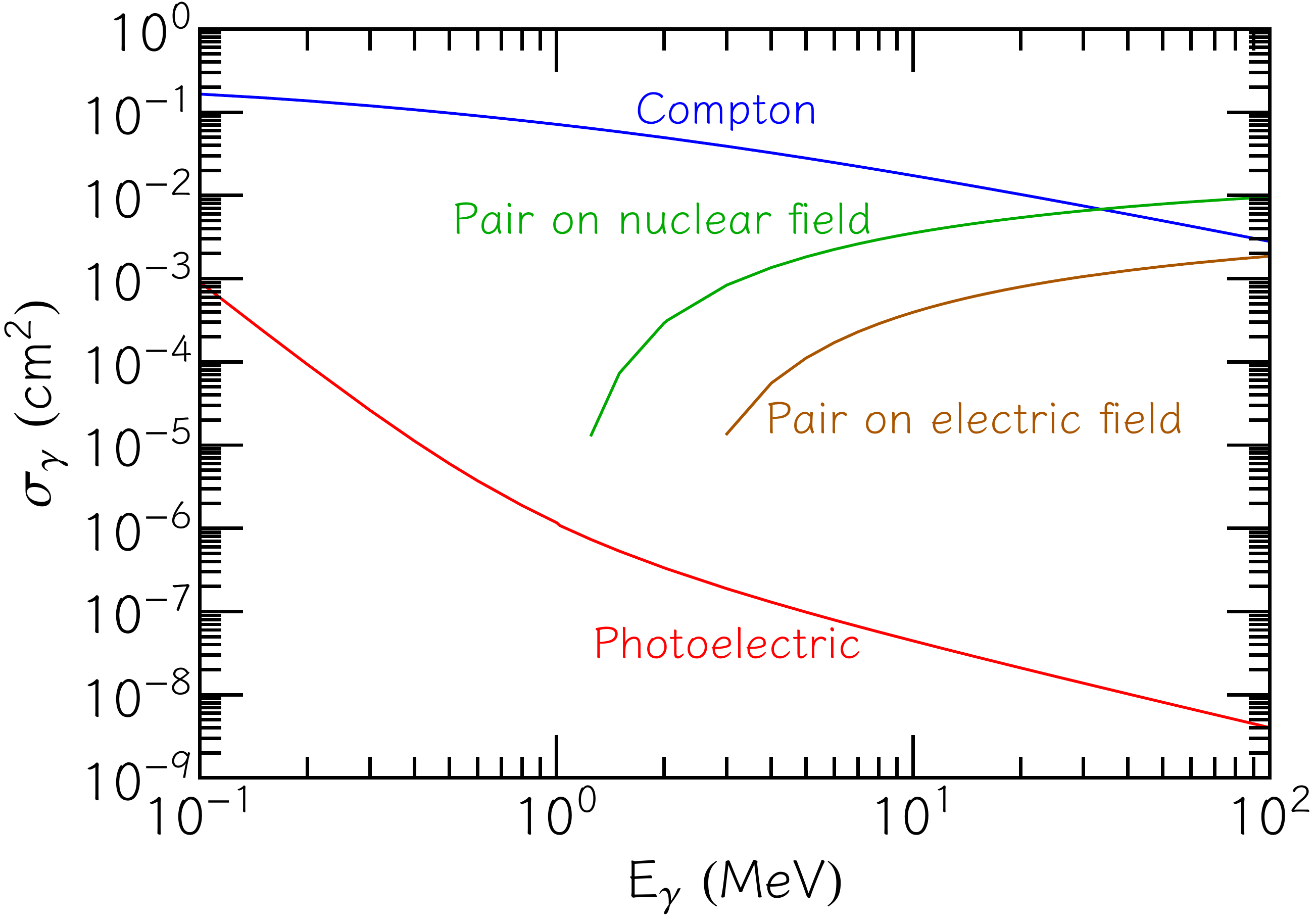}
	\caption{\footnotesize Cross sections for different photon absorption processes in 1 gram of LAB (C$_{19}$H$_{32}$).
    The figure is produced with the XCOM Photon Cross Sections Database~\cite{xcom}.
    }
	\label{fig:Scattering_Processes}
\end{figure}

The integral cross-section for Compton-like scattering $\sigma_{C} $ 
is given by \cite{Donnelly:1978ty,Zhitnitsky:1979cn,Avignone:1988bv}
\begin{eqnarray}
\sigma_{C} = \frac{g_{ae}^{2}\alpha}{8m_e^{2}k_{a}}\left[ \frac{2m_e^{2}(m_e+E_{a})y}{(m_e^{2}+y)^{2}} +
\frac{4m_e(m_{a}^{4}+2m_{a}^{2}m_e^{2}-4m_e^{2}E_{a}^{2})}{y(m_e^{2}+y)} 
+\frac{4m_e^{2}k_{a}^{2}+m_{a}^{4}}{k_{a}y}\ln\frac{m_e+E_{a}+k_{a}}{m_e+E_{a}-k_{a}}\right]\,,
\label{eq:ccsigma}
\end{eqnarray}
where $k_a$ and $E_a=5.49$~MeV are the momenta and the energy of
the axion respectively, $m_e$ is the electron mass and $y = 2\,m_e\,E_a + m_a^2$. 
At fixed value of $g_{ae}$, the phase space contribution to the cross section is approximately independent of the axion mass for $m_a \lesssim 2$~MeV and the integral cross section reduces to
\begin{equation}
    \sigma_{C} \approx g_{ae}^2\times 4.3\times 10^{-25}\,{\rm cm}^2\,.
    \label{eq:ccmassless}
\end{equation}
In the axio-electric effect, which is
analogue of the photo-electric effect, the
axion disappears and an electron is emitted from an atom with an energy equal to the difference between  the absorbed-axion energy and the electron binding energy $E_b$. The cross section for this process is given by 
\begin{equation}
    \sigma_{ae} = \sigma_{pe} \frac{g_{ae}^2}{\beta}\frac{3\,E_a^2}{16\,\pi\,\alpha\,m_e^2}\,\left(1-\frac{\beta^{2/3}}{3}\right)\,,
\end{equation}
where $\beta=|\kappa_a|/E_a$ and $\sigma_{pe}$ is the photoelectric cross section  in  the medium~\cite{xcom}. 
As shown in Fig.~\ref{fig:Scattering_Processes}, in JUNO, which is made of LAB (C$_{19}$H$_{32}$), at energies $\sim O({\rm MeV})$, $\sigma_{pe}$ is more than 5 orders of magnitude lower than the Compton scattering cross section. 
Therefore we neglect this
latter process in our work. 
Note, however, that due to the $Z^5$ dependence of $\sigma_{pe}$, the axio-electric effect is the main axion detection process in detectors with high $Z$ active mass~\cite{LUX:2017glr,Fu:2017lfc,XENON:2020rca}.

An axion may also produce electron-positron pairs in the electric field of nuclei or electrons. 
The relevant cross sections for this process were calculated in Refs.~\cite{Bardeen:1978nq,Zhitnitsky:1979cn,Kim:1982xb,Kim:1984ss}, soon after the axion was introduced, since this seemed a promising detection channel.
Nowadays, the interest in this process has declined. 
In our case, this process is subdominant with respect to Compton, as reflected in the corresponding photon case shown in Fig.~\ref{fig:Scattering_Processes}. 
However, we expect this channel to dominate at higher energies and higher values of $Z$. 
We ignore this channel in the present work and postpone a  detailed analysis of this process to a future project. 

Finally, axions with mass $m_a > 2\,m_e$ can decay into electron-positron pairs, with decay length
\begin{equation}
l_{e}=\frac{\gamma v}{\Gamma_{a\rightarrow e^{+}e^{-}}}  \simeq 
0.33 \,\frac{E_a}{m_a}\frac{\sqrt{1-\frac{m_a^2}{E_a^2}}}{\sqrt{1-\frac{4 m_e^2}{m_a^2}}} 
\left( \frac{g_{ae}}{10^{-11}} \right)^{-2}
\left( \frac{m_{a}}{\rm MeV} \right)^{-1}\, d_{\odot} \,. 
\label{eq:decayel}
\end{equation}
Therefore, the axion flux arriving on Earth is reduced by a factor $\exp(-d_{\odot}/l_e)$ as shown in Eq.~\eqref{eq:phia}.

\subsubsection{Axion-photon coupling}
Axions coupled with photons can be converted into photons in the electric field of charged particles $Ze$ via the inverse Primakoff effect $a+Ze \rightarrow \gamma + Ze$. The differential cross section is given by
\cite{Brdar:2020dpr,Cui:2022owf}
\begin{equation}
    \frac{d \sigma_{\rm{P}}}{d\Omega_a} = \frac{g_{a\gamma}^2\,\alpha}{4\pi} \frac{k_a^4}{q^4}\,\sin^2\theta_a\,F^2(q)\,,
    \label{eq:pcsigma}
\end{equation}
where $\theta_a$ is the scattering angle, $d\Omega_a = d\phi_a d\cos\theta$, and $F(q)$ is the atomic form factor, with $q^2=m_a^2-2\,E_\gamma\,(E_a - k_a\,\cos\theta_a)$ and $E_\gamma \approx E_a$ is the energy of the outgoing photon. 
We use the following atomic form factor, which includes the electron screening of the nuclear charge \cite{Kim:1973he,Tsai:1973py}
\begin{equation}
    F^2(q) = Z^2 \left[\frac{a^2(Z)\,|q^2|}{1+a^2(Z)\,|q^2|}\,\frac{1}{1+|q^2|/d(A)}\right]\,,
\end{equation}
where $a(Z) = 111\,Z^{-1/3}/m_e$ and $d(A) = 0.164\,{\rm GeV}^{2}\,A^{-2/3}$, with $A$ the atomic mass number.

In addition, axions can decay into two photons with decay length
\begin{equation}
l_{\gamma}=\frac{\gamma v}{\Gamma_{a\rightarrow \gamma\gamma}}\simeq
2.64\,\frac{E_a}{m_a}\sqrt{1-\frac{m_a^2}{E_a^2}}
\left( \frac{g_{a\gamma}}{10^{-8}\,{\rm GeV}^{-1}} \right)^{-2}
\left( \frac{m_{a}}{\rm MeV} \right)^{-3}\, d_{\odot}\,.
\label{eq:decayph}
\end{equation}
Therefore the axion flux arriving on Earth is reduced by a factor $\exp(-d_{\odot}/l_\gamma)$ as discussed in Eq.~\eqref{eq:phia}.
Since the decay rate is proportional to $m_a^3$, the decay becomes the dominant process for large values of $g_{a\gamma}^2\,m_a^3$.

\section{Constraining axion couplings}
\label{sec:sensitivity}

\subsection{Likelihood analysis}\label{sec:Numerical}

Here, we outline the fitting procedure that we have adopted to characterize the sensitivity of the JUNO detector. 
JUNO's construction is expected to be completed at the end of 2022~\cite{JUNO:2021vlw}.
We estimate the number of events expected to be detected ($N_{\rm exp}$) using Fig.~11 of Ref.~\cite{JUNO:2020hqc}, which shows the expected event spectra in ten years of data taking, obtained assuming only SM physics.
In order to forecast the detector sensitivity, we define the $\chi^2$ function (see, e.g., Refs.~\cite{Baker:1983tu,JUNO:2020hqc})
\begin{equation}
\begin{split}
\chi^2= & 2\times\sum_{i}\left(  N_{i,\rm pre}-N_{i,\rm exp}+N_{i,\rm exp}\times \log\frac{N_{i,\rm exp}}{N_{i,\rm pre}}\right)  +\left( \frac{\varepsilon_{\rm sb}}{\sigma_{\rm sb}}\right) ^2+\left( \frac{\varepsilon_{\rm rb}}{\sigma_{\rm rb}}\right) ^2, \\
 N_{i,\rm pre} &= (1+\varepsilon_{\rm sb})\times B_{i,\rm sb} +  (1+\varepsilon_{\rm rb})\times B_{i,\rm rb} + \dfrac{S}{\sqrt{2\pi}\bar{\sigma}}\times e^{-\frac{(\bar{E}-E_i)^2}{2\bar{\sigma}^2}},
  \end{split}
\label{equ:spectrumT}
\end{equation}
where $N_{i,\rm exp}$ is the number of solar neutrino events expected to be observed in the $i^{th}$ energy bin, with energy $E_i$~\cite{JUNO:2020hqc}, $N_{i,\rm pre}$ is the predicted number of events in this energy bin assuming the presence of axions, whereas $B_{i,\rm sb}$ and $B_{i,\rm rb}$ represent the solar neutrino and the radioactive background events,\footnote{We are using $N_{i,\rm exp} = B_{i,\rm sb} + B_{i,\rm rb}$.} taken from Ref.~\cite{JUNO:2020hqc}. 
Here, $ \varepsilon_{sb} $ and $ \varepsilon_{rb} $ are the nuisance parameters and the corresponding solar and radioactive background normalization uncertainties are given by $ \sigma_{sb}  $ and $ \sigma_{rb} $, respectively.

\begin{figure}[!t]
\includegraphics[width=0.62\textwidth]{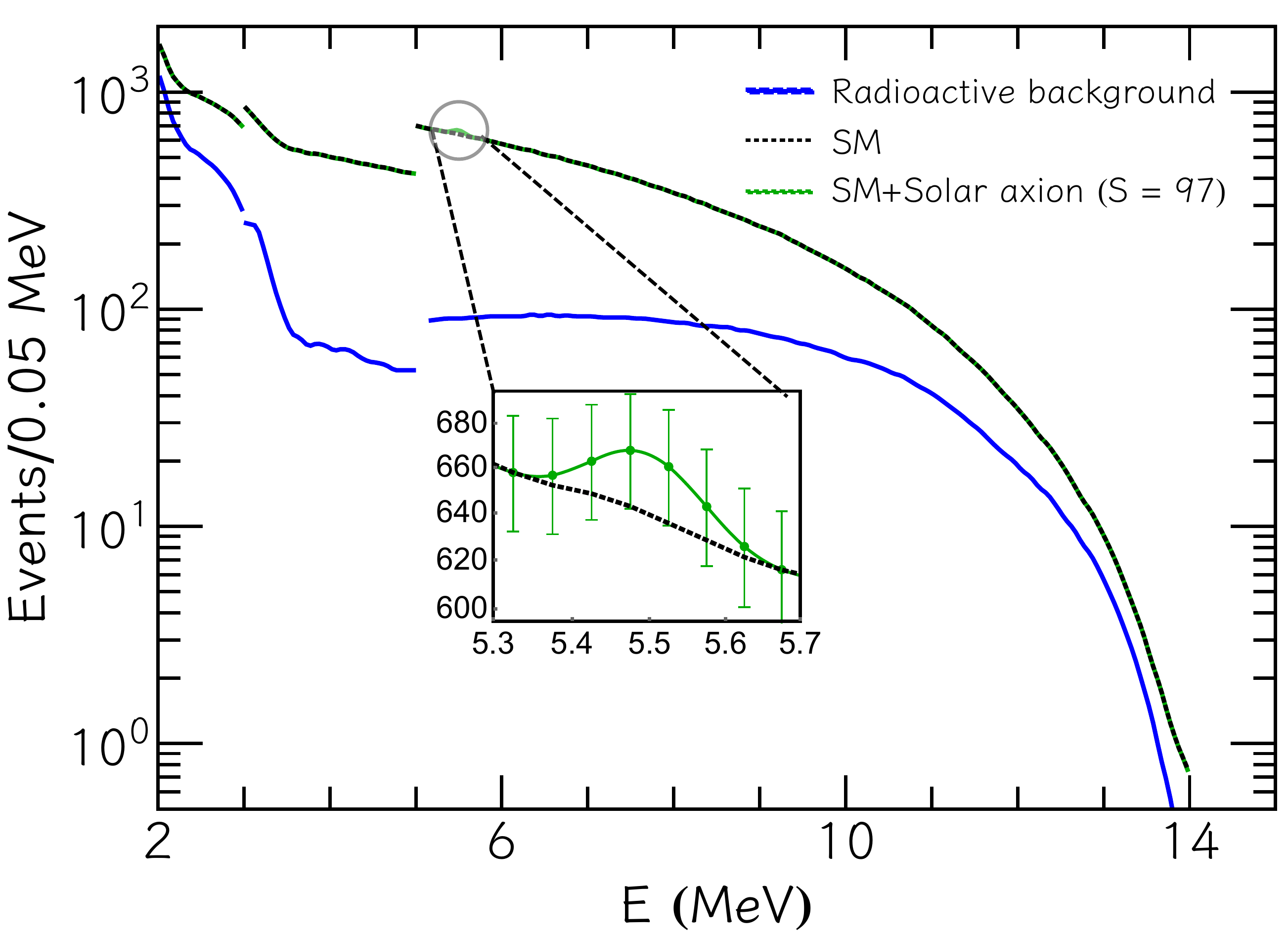}
\caption{\footnotesize  Expected events spectra for JUNO in 10 years of data taking. The solid blue line represents the radioactive background spectrum. The dotted-black curve shows the standard model (SM) spectrum, obtained summing the expected solar neutrino and the radioactive background events~\cite{JUNO:2020hqc}. The solid green curve represents the spectrum expected to be detected in presence of solar axions, with a peak intensity $S=97$ counts in 10 years, corresponding to the $90\%$ C.L. sensitivity. 
The in-set shows an enlarged picture of the axion induced 5.49 MeV bump. Error bars represent the statistical Poissonian errors.
}
\label{fig:JunoEvents}
\end{figure}
The new physics contribution has been modeled as a Gaussian function, with $S$ parametrizing the expected axion peak intensity, centered at $\bar{E}=5.49$~MeV and with a width $\bar{\sigma} = 0.07$~MeV, given by the
detector energy resolution in Eq.~\eqref{eq:eres} evaluated at $5.49$~MeV.  Fig.~\ref{fig:JunoEvents} displays with a dotted black line the total number of expected events from SM in bins with a width of $0.05$ MeV, while the blue line represents the contribution from radioactive background only. The breaks in the spectra at 3 and 5 MeV are related to the energy-dependent FV discussed above. 
On the other hand, the green line shows the total events expected to be detected in presence of solar axions, for a representative value $S=97$ counts in ten years, corresponding to the $90\%$ confidence level (C.L.) sensitivity, as discussed in the following. 
The axion bump at 5.49 MeV can be observed. 

\begin{figure}[!t]
\includegraphics[width=0.62\textwidth]{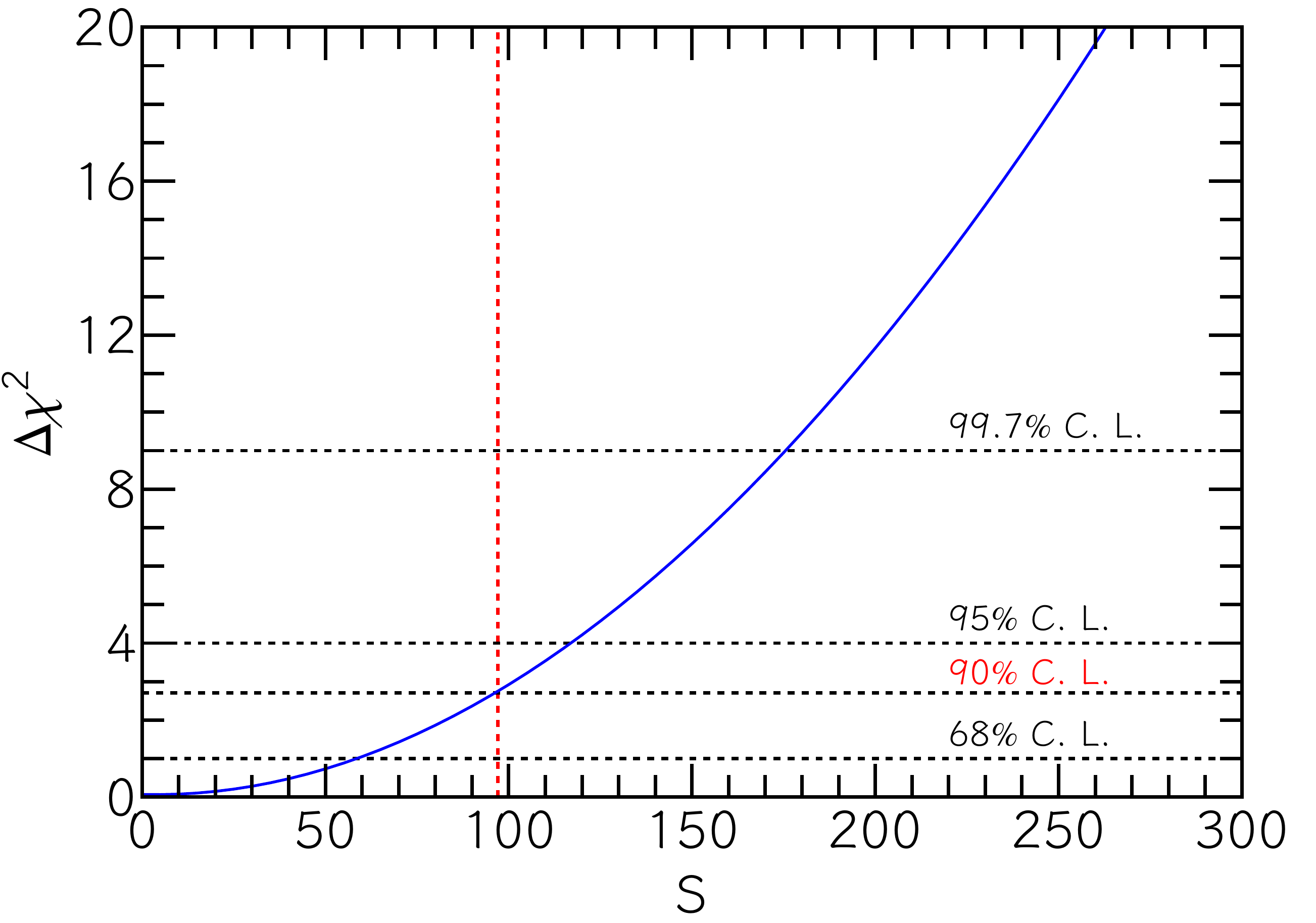}
\caption{\footnotesize $ \Delta \chi^{2} $ as a function of the peak intensity $S$.
Here, the horizontal dotted-black lines represent different significance levels. In our analysis, we forecast the sensititivity at $90\%$ C.L., which corresponds to $S_{\rm lim} = 97$, indicated by the vertical dotted-red line.
}  

\label{fig:ChiSq}
\end{figure}

To perform a $ \chi^{2} $ test, we marginalize over the nuisance parameters and fix the normalization uncertainties for solar and radioactive background as $\sigma_{sb} = 5\%$ and $\sigma_{rb} =15\%$, respectively. By construction, the $\chi^2$ function is minimized for $S=0$ (no axion events).
A plot of $\Delta\chi^2=\chi^2(S)-\chi_{{\rm min}}^2$ as a function of the peak intensity $S$ is shown in Fig.~\ref{fig:ChiSq}.  
By fixing $\Delta \chi^{2}(S) = 2.71$, we find that the JUNO sensitivity at $90\%$ C.L.\footnote{We choose to forecast the sensitivity at $90\%$ C.L. to make a direct comparison with Borexino limits~\cite{Borexino:2012guz}.} is $S_{\rm lim} = 97$ counts in 10 years.
From Eq.~\eqref{eq:nev}, this upper limit  can be used to constrain the product of the axion flux $\Phi_a$ with 
the cross section of processes having as targets electrons $\sigma_{a-e}$ or Carbon nuclei $\sigma_{a-C}$ via~\cite{Borexino:2012guz} 
\begin{equation}
S_{\rm events} = \Phi_{a}\,\sigma_{a-e,C}\,N_{e,C}\,T\,\varepsilon \leq S_{\rm lim} \;,
\label{eq:Events}
\end{equation}
where $N_{e}\simeq 5.5 \times 10^{33}$ and $N_{C}\simeq 7.1 \times 10^{32}$ are the numbers of electrons and carbon nuclei in the 16.2 kton FV, respectively, $T=10$~years is the measurement time and $\varepsilon=1$ is
the detection efficiency.\footnote{Since we used the fiducial volume, rather than the total volume, the detection efficiency is considered as one.} 
Therefore, the individual rate limits at $90\%$ C.L. are
\begin{eqnarray}
\Phi_{a}\sigma_{a-e} \leq 5.6\times 10^{-41} \rm{s^{-1}}\,,\\  \label{SNU_e} \Phi_{a}\sigma_{a-C} \leq 4.3\times 10^{-40}
 \rm{s^{-1}}\,,\\\label{SNU_C}
\end{eqnarray}
almost two order of magnitude smaller than
the corresponding Borexino limits~\cite{Borexino:2012guz}. These values describe the sensitivity limit to a model-independent value $\Phi_{a}\sigma_{a}$. 
In this framework, electrons are targets for the Compton effect, while Carbon nuclei for the inverse Primakoff process.

Analogously, in the case of axion decays into photons or electron-positron pairs inside the detector, limits can be obtained by requiring
\begin{equation}
    S_{\rm dec} = \Phi_a \frac{V}{l_i} \varepsilon T\leq S_{\rm lim}\,.
    \label{eq:Sdec}
\end{equation}

To conclude this section, we point out that in general the value of the position $\bar{E}$ and dispersion $\sigma$ of the Gaussian signal in Eq.~\eqref{equ:spectrumT} could be different for different interactions or decay processes, as discussed by the Borexino collaboration in Ref.~\cite{Borexino:2012guz}. This implies a different value of $S_{\rm lim}$ for each process. In absence of a dedicated Monte Carlo simulation of JUNO response, for simplicity, throughout this work we adopt a unique value of
$S_{\rm lim}=97$ counts in 10 years to serve our purpose. 
In the next sections, we present our sensitivity study and results derived from the assumptions above.
%

\subsection{Joint Sensitivity on $(g_{ae}$,~$g_{3aN})$} 
\begin{figure}[!t]
\includegraphics[width=0.62\textwidth]{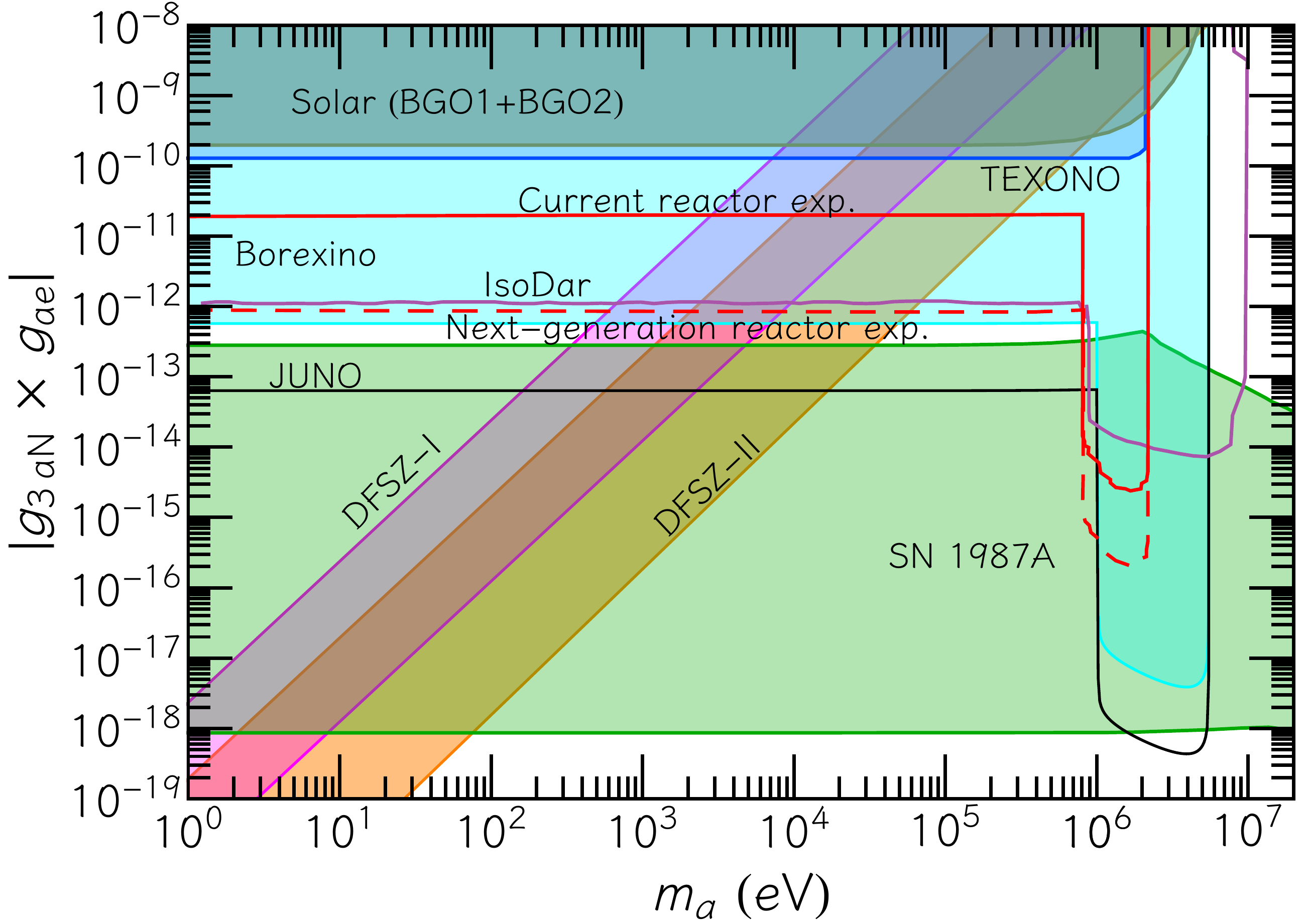}
\caption{\footnotesize  Exclusion region plot in the ($|g_{3aN} \times g_{ae}|$, $ m_a $) plane at 90\% C.L. 
The solid black line represents the JUNO sensitivity. Details on the other constraints are given in the main text.
}

\label{fig:gAegA3NVaryMa}
\end{figure}

From Eq.~\eqref{eq:Events}, the expected number of events due to Compton conversion in the FV is given by
\begin{equation}\label{eq:Counts_CC}
S_{C} = \Phi_{a}\sigma_{C}N_{e}T \;,
\end{equation}
where $\sigma_{C}$ is the Compton conversion cross sections in Eq.~\eqref{eq:ccsigma}.
%
The axion flux  is proportional to $ g^{2}_{3aN} $ (see Eq.~\eqref{eq:phia}), whereas the cross section $\sigma_{C}$ for $ m_a \lesssim 2 $ MeV can be found in Eq.~\eqref{eq:ccmassless}.
Since $ (k_a/k_{\gamma})^3 \simeq 1$ for $m_a \lesssim 1$ MeV in Eq.~\eqref{eq:phia}, we can simplify Eq.~\eqref{eq:Counts_CC} to 
\begin{equation}
S_{C} = g^{2}_{3aN} \times  g^2_{ae} \times 2.42 \times 10^{28} \;.
\label{eq:Bounds_CC}
\end{equation} 
Therefore, at $90\,\%$ C.L.the sensitivity on the product $| g_{3aN} \times g_{ae} |$ is
\begin{equation}
| g_{3aN} \times g_{ae} | \leq 6.33 \times 10^{-14}\qquad {\rm for}\,\,m_a\lesssim 1~{\rm MeV}\;.
\label{eq:Limits_CC}
\end{equation} 
As shown in Fig.~\ref{fig:gAegA3NVaryMa}, this result is one order of magnitude stronger than the Borexino bound $ | g_{3aN} \times g_{ae} | \leq 5.5 \times 10^{-13} $~\cite{Borexino:2012guz} (cyan region). For larger values of the mass, $|g_{a3N} \times g_{ae}|$ depends on $m_a$ due to the kinematic factors in Eqs.~\eqref{eq:phia} and \eqref{eq:ccsigma}.

In addition, for $m_a>2\,m_e$ axions can decay into electron-positron pairs. 
The number of events expected to be detected by JUNO through  decay into electron positron pairs is given by
\begin{equation}\label{eq:S-ep-Decay}
S_{ e^+  e^-} = N_{ e^+ e^-} T  \;,
\end{equation}
where
\begin{equation}\label{eq:Number-ep-Decay}
N_{ e^+ e^-} = \Phi_a\,\frac{V}{l_{e}}    \,,
\end{equation}
is the number of $a \rightarrow  e^+ e^-$ decays in the detector, 
with  $V$ indicating JUNO fiducial volume~\cite{JUNO:2015zny} and $l_e$ the decay length in Eq.~\eqref{eq:decayel}. 

Fig.~\ref{fig:gAegA3NVaryMa} shows the parameter space that can be explored by JUNO through the processes mentioned above. 
At sufficiently small axion-electron coupling  ($g_{ae}\lesssim 10^{-11}-10^{-12}$), we can ignore the reduction in the flux in Eq.~\eqref{eq:phia} due to the $e^{-d_{\odot}/l_{\rm e}}$ term.
In this case, JUNO would be able to  probe the region  $|g_{3aN}\times g_{ae}|\sim O(10^{-18})$ for $1~{\rm MeV}\lesssim m_a \lesssim 5.5~{\rm MeV}$. 
Also in this case, the JUNO sensitivity is an order of magnitude stronger than the Borexino bound.\footnote{Notice that the number of axion decays into electron positron pairs was not considered in~\cite{Borexino:2012guz}. For this reason, here we estimate the Borexino bound through Eq.~\eqref{eq:Sdec} using as benchmark $S_{\rm{lim}}=6.9$ counts in 536 days (see Table I in \cite{Borexino:2012guz}) and the FV of Borexino $\sim 1.15\times 10^{8}$~cm$^{3}$.} 

In Fig.~\ref{fig:gAegA3NVaryMa}, we also show other bounds and the sensitivity of future experiments. 
For $m_a \lesssim 1$~MeV the region $|g_{3aN}\times g_{ae}|\gtrsim 2\times 10^{-10}$ is excluded at $90\%$ C.L. due to the non-observation of events induced by solar 5.49 MeV axions through axio-electric effect in Bi$_4$Ge$_3$O$_{12}$ (BGO) bolometric detectors~\cite{Derbin:2013zba,Derbin:2014xzr} (see the brown region in Fig.~\ref{fig:gAegA3NVaryMa}). The TEXONO collaboration~\cite{TEXONO:2006spf} (blue region) excludes $|g_{3aN} \times g_{ae}|\gtrsim 1.3\times 10^{-10}$ at $90\%$ C.L. for $m_a \lesssim 10^{6}$~eV from the non-observation of axions produced in nuclear transition and detectable after Compton effect in a high-purity germanium detector. 
Current reactor experiments (solid red line)~\cite{AristizabalSierra:2020rom} reach a sensitivity $| g_{3aN}\times g_{ae}|\sim 10^{-11}$, while next-generation experiments (dashed red line)~\cite{AristizabalSierra:2020rom} could compete with the Borexino limits. 
A similar sensitivity (see purple line) will be reached by the Isotope-Decay-at-Rest (IsoDar) experiment, searching for axions using
monoenergetic nuclear de-excitation photons from a beam dump~\cite{Waites:2022tov}.
For $m_a<14.4$~keV, experiments searching for solar $^{57}$Fe axions detectable through axio-electric
absorption in dark matter detectors using Germanium, such as EDELWEISS III~\cite{EDELWEISS:2018tde},
CDEX~\cite{CDEX:2016rpr} and MAJORANA DEMONSTRATOR~\cite{Majorana:2016hop}, or Xenon targets, such as PANDAX-II~\cite{PandaX:2017ock}, constrain the combination $g_{aN}^{\rm eff}=|-1.19\,g_{0aN} + g_{3aN}|$. Therefore, assuming $g_{0aN}\approx 0$, these experiments would exclude at most $g_{3aN}\gtrsim 10^{-17}.$
These bounds are not shown in Fig.~\ref{fig:gAegA3NVaryMa} since they cannot be translated univocally into a bound on $|g_{3aN} \times g_{ae}|$. 
The supernova (SN) bound from the cooling of SN 1987A is the strongest constraint in this region of the parameter space and it is obtained multiplying the values of the constraints on the individual couplings, i.e. $9.1\times 10^{-10} \lesssim g_{3aN} \lesssim 10^{-6}$ (see Appendix~\ref{app:supernova} and Ref.~\cite{Carenza:2019pxu}) and $10^{-9} \lesssim g_{ae} \lesssim 3\times 10^{-7}$ for $m_a \lesssim 1$~MeV. 
We observe that JUNO would probe the  region of the parameter space for $m_a\lesssim 1$~MeV and $| g_{3aN}\times g_{ae}|\sim 5\times 10^{-13}$, currently unexplored by direct detection experiments.
Finally, we have also displayed the allowed parameter space for the DFSZ-I and DFSZ-II axion models using the light magenta and light orange regions, respectively~\cite{Dine:1981rt,Zhitnitsky:1980tq,AristizabalSierra:2020rom}. 
%


\begin{figure}[!t]
\includegraphics[width=0.48\textwidth]{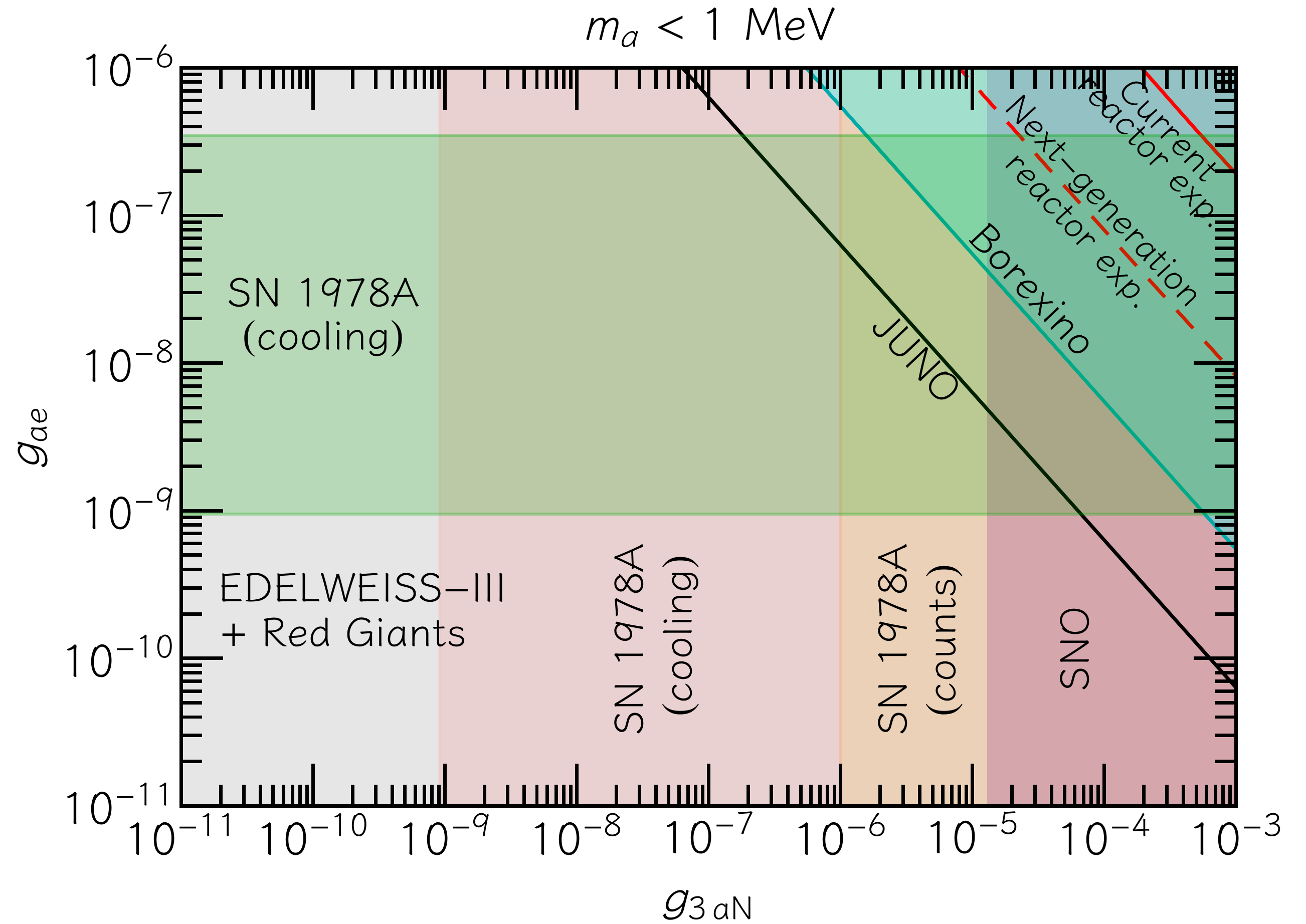}
\,\,\,
\includegraphics[width=0.48\textwidth]{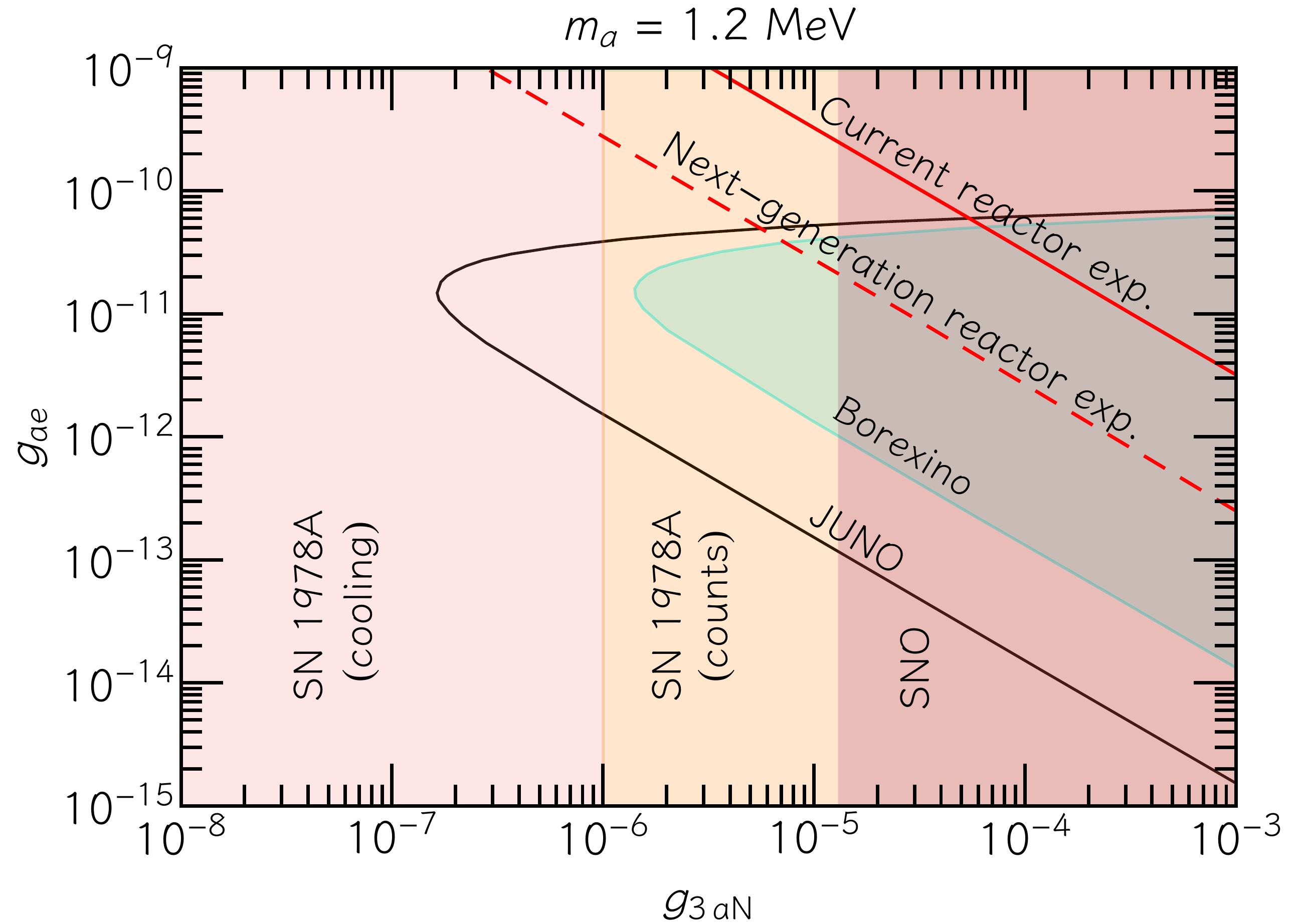}
\caption{\footnotesize 
\textit{Left panel:} Exclusion region plot in the ($g_{ae}$, $g_{3aN}$) plane at 90\% C.L. for $m_a<1$~MeV. JUNO sensitivity is shown using the solid black line. Similar bound for Borexino is shown using the solid cyan and colored region. Red lines represent sensitivities of reactor experiments as examined in \cite{AristizabalSierra:2020rom}. Details on the astrophysical bounds are given in the text.
\textit{Right panel:} same as left panel but for $m_a$ = 1.2 MeV.}
\label{fig:gAegA3N}
\end{figure}

In Fig.~\ref{fig:gAegA3N}, we show exclusion region plots in the ($g_{ae}$, $g_{3aN}$) plane, at fixed value of the axion mass. In the left panel, the solid black line represents the JUNO sensitivity for axion mass $m_a < 1$~MeV, obtained using Eq.~\eqref{eq:Limits_CC}. For comparison, we show also bounds arising from the Borexino detector (cyan-colored region) and the sensitivities of current (solid red line) and next-generation neutrino reactor experiments (dashed red line) \cite{AristizabalSierra:2020rom}.
It can be noticed that even in this case JUNO 
has the potential to set constraints on axion couplings that are almost an order of magnitude tighter than those derived from the previous Borexino analysis. 
However, this region of the parameter space is also constrained by EDELWEISS-III~\cite{EDELWEISS:2018tde} and astrophysical arguments.
In particular, the red giant (RG) bound excludes $g_{ae}\gtrsim 1.6 \times 10^{-13}$~\cite{Straniero:2020iyi,Capozzi:2020cbu}. 
In both panels of Fig.~\ref{fig:gAegA3N} we show the SN cooling bound on $g_{ae}$~\cite{Lucente:2021hbp,Ferreira:2022xlw} (green region) and $g_{3aN}$~\cite{Carenza:2019pxu} (lighter purple) and constraints arising from additional event counts at
Kamiokande-II~\cite{Engel:1990zd} (lighter orange) and from the SNO analysis~\cite{Bhusal:2020bvx} (purple). 
Notice that to express different sensitivities and bounds, we adopts the same color codes throughout the work.

In the right panel of Fig.~\ref{fig:gAegA3N}, we show JUNO sensitivity for $m_{a}=1.2$~MeV, where the axion decay into electron-positron pairs is relevant. Using Eq.~\eqref{eq:Number-ep-Decay} for $m_a = 1.2 $ MeV, and $T$ in Eq.~\eqref{eq:S-ep-Decay},  limits on the axion couplings for the JUNO detector  can be calculated as 
\begin{equation}
| g_{3aN} \times g_{ae} | \leq  \dfrac{  1.52 \times 10^{-18}  }{\sqrt{\exp(- 4.25 \times 10^{21}  \, g^2_{ae} )}}\;.
\label{eq:Limits_ep-Decay}
\end{equation} 
We derive similar limits for the Borexino detector,\footnote{Events from axion decays into electron-positron pairs were neglected in Ref.~\cite{Borexino:2012guz}} obtaining
\begin{equation}
| g_{3aN} \times g_{ae} | \leq  \dfrac{  1.32 \times 10^{-17} }{\sqrt{\exp(- 4.25 \times 10^{21} \, g^2_{ae} )}}\;.
\label{eq:LimitsBor_ep-Decay}
\end{equation} 
%
In this mass range, the sensitivity has a nose-like shape, since for couplings smaller than the lower limit not enough axions decay inside the detector, while for values larger than the upper limit, axions decay before reaching the Earth. Also in this case, JUNO is the experiment with the best sensitivity. 
This region of the parameter space is not constrained by red giants, since the axion production is Boltzmann suppressed for $m_{a}\gtrsim O(10)$~keV.
Thus, the only competitive bound in this region is the SN limit.

\subsection{Joint Sensitivity on $g_{3aN}$,  $g_{a\gamma}$ }

\begin{figure}[!t]
\includegraphics[width=0.62\textwidth]{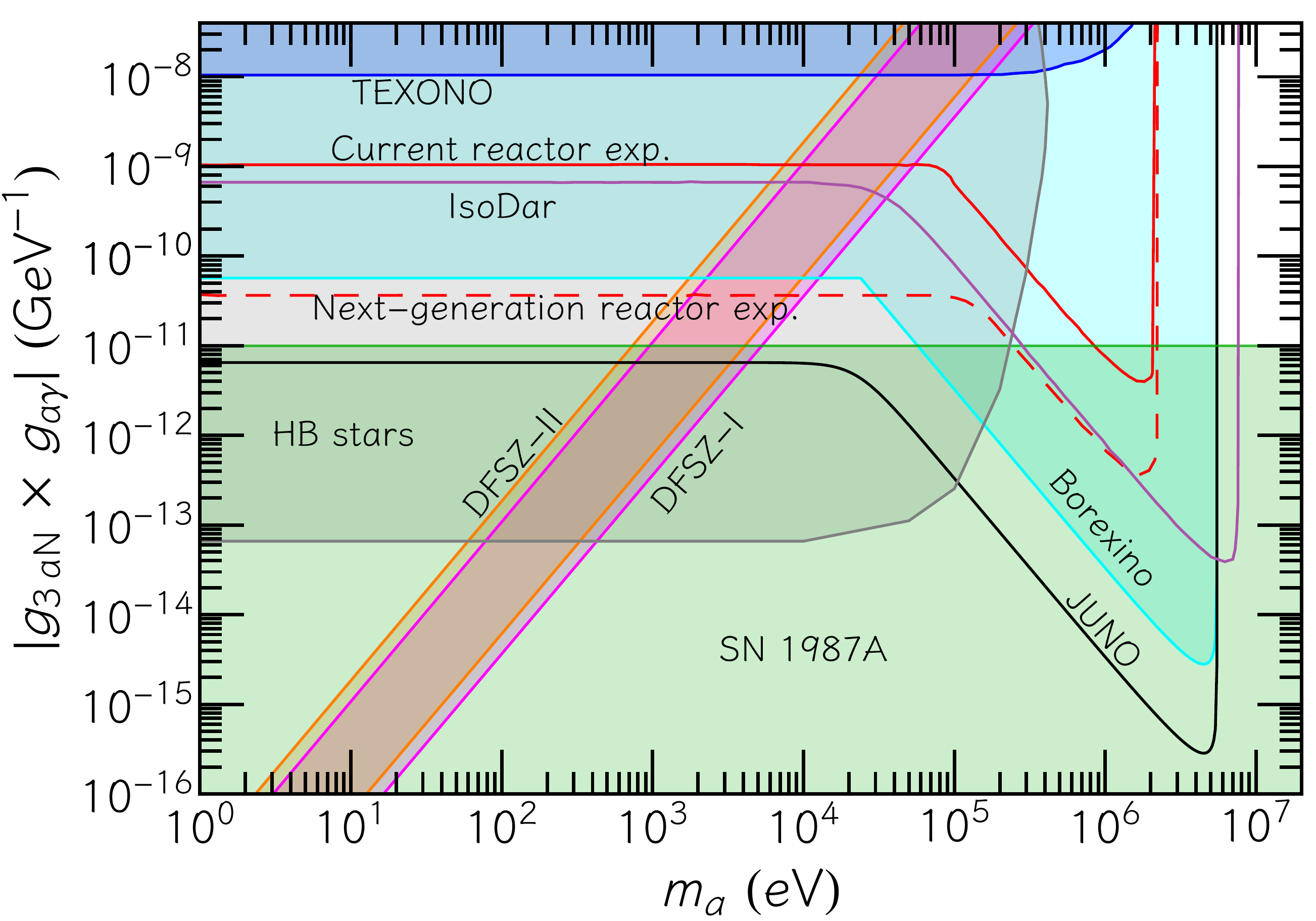}
\caption{\footnotesize  Exclusion region plot in the ($|g_{3aN} \times g_{a\gamma}|$, $m_a$) plane at 90\% C.L. The solid black line represents JUNO sensitivity. Details on the other constraints and sensitivities are given in the main text.
  }
\label{fig:gagprodma}
\end{figure}

Axion coupled to photons may be detected in JUNO through the Primakoff process or through axion decay into two photons. 
The JUNO sensitivity in this case is shown in Fig.~\ref{fig:gagprodma}.

The number of expected events due to inverse Primakoff conversion is given by
\begin{equation}
S_{P} = \Phi_a\sigma_{P}N_{C}T\varepsilon_{P}\;, \label{Counts_PC}
\end{equation}
where $N_C$ is the number of Carbon nuclei in the FV and $\sigma_{P}$ is the Primakoff conversion cross section obtained integrating Eq.~\eqref{eq:pcsigma} over the scattering angle. 
In the small mass limit ($m_a\lesssim 10$~keV) and  under the assumption that $\Phi_a = \Phi_{a0}$ (i.e., $m_a^2 ({\rm eV}) \times g_{a\gamma} ({\rm GeV}^{-1}) < 1.2\times 10^{4}\,\,{\rm eV}^2\,\,{\rm GeV}^{-1}$), 
the JUNO sensitivity reaches at $90~\%$ C.L.
\begin{equation}
| g_{3aN} \times g_{a\gamma} | \lesssim 6.5 \times 10^{-12}\,{\rm GeV}^{-1}\qquad {\rm for}\,\,m_a\lesssim 10~{\rm keV}\;,
\label{eq:Limits_PC}
\end{equation} 
improving on the Borexino limits~\cite{Borexino:2012guz} by almost one order of magnitude, as shown in Fig.~\ref{fig:gagprodma}. 
For larger values of the mass, the axion decay becomes important and the sensitivity on $| g_{3aN} \times g_{a\gamma} |$ depends on $m_a$. 
Indeed, the number of events expected to be detected by JUNO after axion decays into two photons is given by 
\begin{equation}\label{Counts_2G}
S_{2\gamma} = N_{\gamma}T \;, 
\end{equation}
where $T$ is the exposure time and $N_\gamma$ is the number of decays inside the detector
\begin{equation}
N_\gamma = \Phi_a\,\frac{V}{l_{\gamma}}\,,
\end{equation}
with $\Phi_a$ in Eq.~\eqref{eq:phia} and $l_{\gamma}$ in Eq.~\eqref{eq:decayph}. 
Assuming ultrarelativistic axions, $\beta \sim 1$, the axion decay implies a limit at $90 \%$ C.L.
\begin{equation}
  | g_{3aN} \times g_{a\gamma} |\times m_a^2 \lesssim 3.3 \times 10^{-12}\,\,{\rm eV}\,,
\end{equation}
for $10~{\rm keV}\lesssim m_a < 5$~MeV, as shown in Fig.~\ref{fig:gagprodma}.
In this case, JUNO is capable of exploring axion couplings  $| g_{3aN} \times g_{a\gamma} | \sim \mathcal{O} (10^{-15}) $ GeV$ ^{-1} $ for $m_a\sim \mathcal{O} ({\rm MeV})$. 
For axion masses closer to the limit of 5.49 MeV, the dependence of the bound on the axion mass changes since the ultrarelativistic assumption for the axions becomes invalid. 

For comparison, in Fig.~\ref{fig:gagprodma} we show also the Borexino bound (cyan) as well as sensitivities of the current (next-generation) neutrino reactor experiments~\cite{AristizabalSierra:2020rom} in solid (dashed) red lines and of the IsoDar experiment~\cite{Waites:2022tov} in purple.
Furthermore, we show the TEXONO bound~\cite{TEXONO:2006spf} (blue region), constraining $|g_{3aN} \times g_{a\gamma}|\lesssim 7.7\times 10^{-9}$~GeV$^{-1}$ at $90\%$ C.L. for $m_a \lesssim 10^{5}$~eV from the non-observation of axions produced in nuclear transition and detectable after Primakoff conversion in the detector.
Finally, we  show astrophysical bounds from  HB stars and from SN.  
The SN 1987A bound (green region) is obtained from the constraints on the individual couplings $9.1\times 10^{-10} \lesssim g_{3aN} \lesssim 10^{-6}$~\cite{Carenza:2019pxu} and $7\times 10^{-9}~{\rm GeV}^{-1} \lesssim g_{a\gamma} \lesssim 2\times 10^{-6}~{\rm GeV}^{-1}$ for $m_a\lesssim 10$~MeV~\cite{Caputo:2021rux}. 
The gray region represents the bound from horizontal-branch (HB) stars in globular clusters~\cite{Carenza:2020zil,Lucente:2022wai}. 
Since there is not a HB bound on $g_{3aN}$, we estimate the constraint on the product $|g_{3aN} \times g_{a\gamma}|$ by requiring that $g_{3aN}\lesssim 10^{-3}$, to allow axions to escape from the Sun. 
Our analysis shows that, even for the  $| g_{3aN} \times g_{a\gamma} |$ combination of couplings,
JUNO has the best experimental sensitivity for all the axion masses, improving on the Borexino limit by approximately
one order of magnitude.
Thus, JUNO has the potential of exploring regions of the axion parameter space currently accessible only through astrophysical arguments.

\begin{figure}[!t]
\includegraphics[width=0.48\textwidth]{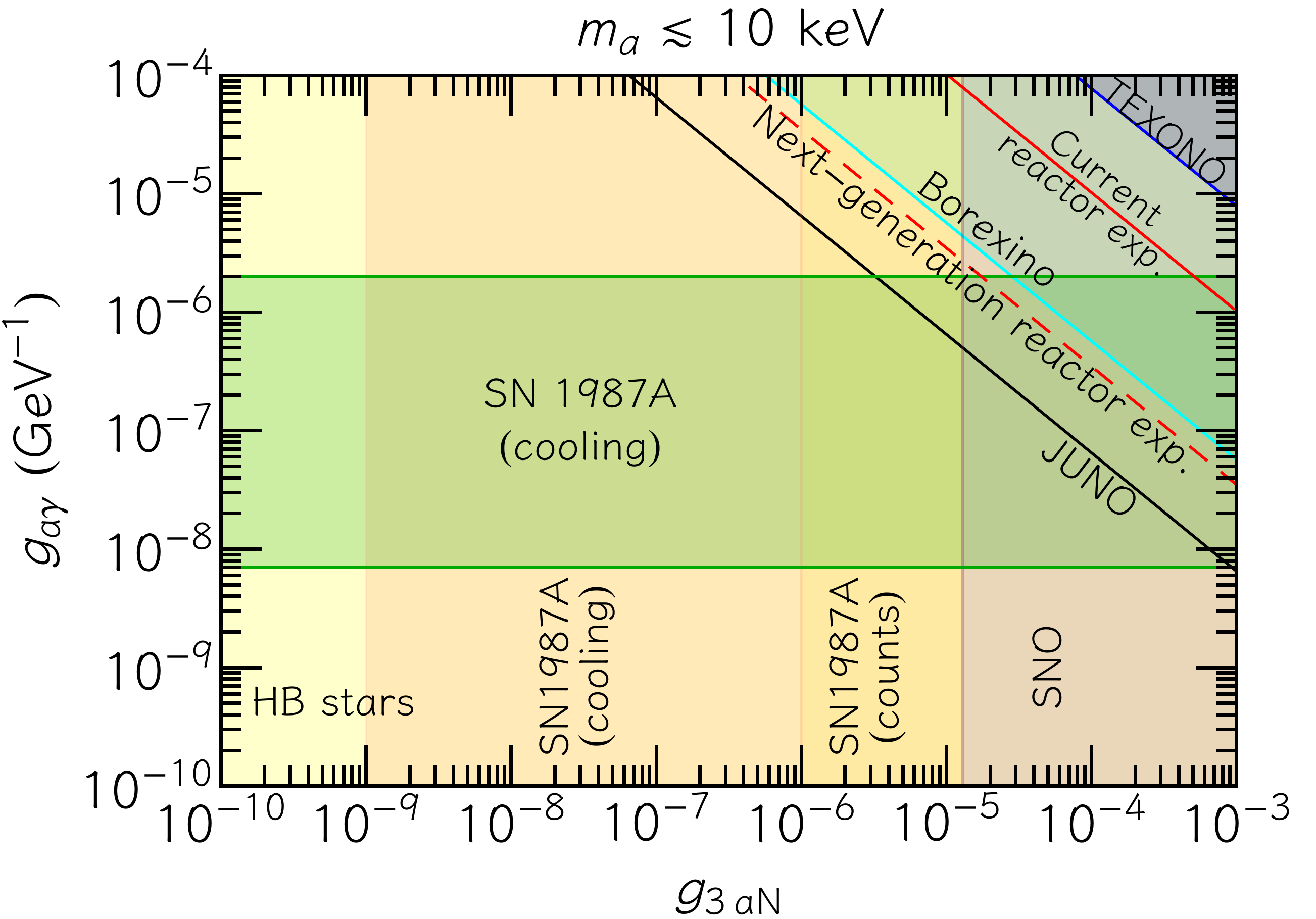}
\,
\includegraphics[width=0.48\textwidth]{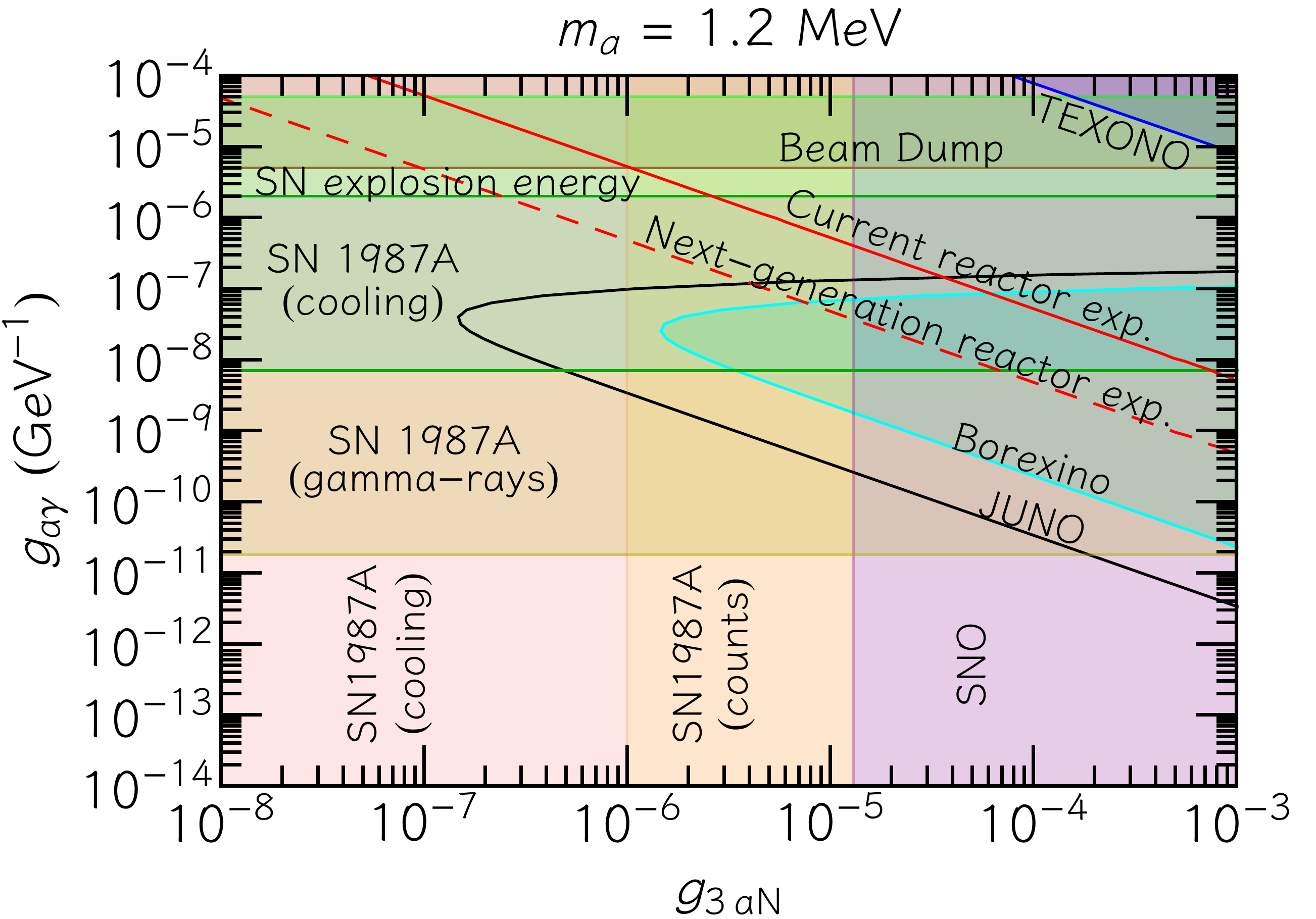}
\caption{\footnotesize 
\textit{Left panel:} Exclusion region plot in the ($g_{a\gamma}$, $g_{3aN}$) plane at $90\%$ C.L. for $m_a\lesssim10$~keV. JUNO sensitivity is shown using the solid black line. The solid cyan region is the Borexino bound. Red lines represent sensitivities of reactor experiments as examined in \cite{AristizabalSierra:2020rom}. Details on the other experimental and astrophysical bounds are given in the text.
\textit{Right panel:} same as left panel but for $m_a$ = 1.2 MeV.}
\label{fig:gagmev}
\end{figure}

Finally, in Fig.~\ref{fig:gagmev} we show the sensitivity in the ($g_{a\gamma}$, $g_{3aN}$) plane at fixed values of the axion mass.
In the left panel, we show the small mass limit case $m_a < 10$~keV, 
where the dominant process is the inverse Primakoff, and in the right panel we show the case of $m_a= 1.2$~MeV, where the dominant process is the axion decay. 
In the small mass limit, the bound follows Eq.~\eqref{eq:Limits_PC} and it improves on all the other experimental bounds and sensitivities. This region is constrained by the SN limits previously discussed and the HB bound~\cite{Ayala:2014pea,Straniero:2015nvc} on $g_{a\gamma}$, which completely excludes the experimental region of interest in this mass range.
On the other hand, at $m_a=1.2$~MeV the JUNO sensitivity has the typical nose-like shape discussed in the previous section and can probe $g_{a\gamma}\sim 10^{-12}$~GeV$^{-1}$ for $g_{3aN}\sim 10^{-3}$. 
This region is not constrained by the HB bound, since the axion production is suppressed for $m_a\gtrsim O(100)$~keV. 
There are, however, other astrophysical and experimental limits. 
Besides the bounds discussed above,
in this mass range the couplings $g_{a\gamma}\gtrsim 5\times 10^{-6}$~GeV$^{-1}$ are excluded by beam dump experiments (brown)~\cite{Agrawal:2021dbo,Dolan:2017osp,Dobrich:2019dxc}, while lower values of the coupling are constrained by requiring that axion decays must not lead to an excessive SN explosion energy (light green)~\cite{Caputo:2022mah} and from the non-observation of a gamma-ray flux in association with the SN 1987A explosion (darker yellow)~\cite{Jaeckel:2017tud,Caputo:2021rux}.

\section{Comparison with future neutrino experiments}
\label{sec:HK}
In this section we evaluate the sensitivity of other forthcoming neutrino  experiments  to detect 5.49 MeV solar axions.
The next-generation Hyper-Kamiokande (HK) neutrino observatory is planned to be installed near Kamioka, in Japan, and is expected to start in 2027~\cite{Hyper-Kamiokande:2018ofw}. 
The HK collaboration plans to use a water-Cherenkov detector with 374 kton fiducial volume with 
energy resolution 
\begin{equation}
\sigma/{\rm MeV} = 0.6 \sqrt{E/{\rm MeV}}\,.
\label{eq:ResHK}
\end{equation}
%
To derive the HK sensitivity on  axion couplings, here we adopt the same procedure described in Sec.~\ref{sec:Numerical}.
We have first calculated the expected solar neutrino events ($N_{i,\,\rm exp}$) for the HK detector. 

We compute the expected number  of neutrinos for the $i^{th}$ energy bin as
\begin{equation}\label{eq:EventsHK}
N_{i,\,\rm exp} =  \int_{E_i}^{E_{i+1}} \frac{dN_{\rm exp}}{d E_{\rm vis}} \,dE_{\rm vis} \;, 
\end{equation}
where, 

\begin{equation}
\frac{dN_{\rm exp}}{d E_{\rm vis}} = N_e \, T \, \varepsilon\, \theta(E_{vis}-E_T)\int dT_{e}  \mathcal{R}(E_{vis}, T_e) \int dE_{\nu}  \frac{d\Phi_{\rm sol}}{dE_\nu} \frac{d\sigma}{d T_{e}} (E_{\nu}, T_e)  \;,
\end{equation}
and $\theta(E_{vis}-E_T)$ is the Heaviside step function.
Here, $N_{e}\simeq 1.5 \times 10^{35}$ represents the number of electrons corresponding to 374 kton detector, $T$ is the 10-year run-time~\cite{Hyper-Kamiokande:2018ofw}, while the detection efficiency $\varepsilon=1$ has been adopted  for the HK detector.
Also, $E_{\nu} $ is the neutrino energy, $T_e$ is the kinetic energy of the recoil electron, and $E_T = 3.5$ MeV \cite{HKresolution:2019} is the threshold energy necessary  to produce an electron. $\mathcal{R}$ is the Gaussian energy resolution  function, having a width given by Eq. \eqref{eq:ResHK}.
%
Using the solar neutrino flux in Ref.~\cite{Bahcall:1996qv} and the differential cross-sections for the neutrino-electron elastic scattering processes in Ref.~\cite{Giunti:2007ry}, Eq. \eqref{eq:EventsHK} predicts $\sim 3 \times 10^7$ events in the HK detector. 
Here, we have only considered the solar background normalization uncertainties, $ \sigma_{sb} = 5\%$.
 %
The $ \chi^{2} $  analysis for the HK detector leads to $S_{\rm lim} = 9900$ at $90\,\%$ C.L.
\begin{equation}
| g_{3aN} \times g_{ae} | \leq 1.22 \times 10^{-13} \qquad {\rm for}\,\,m_a\lesssim 1~{\rm MeV}\;,
\end{equation} 
at $90\,\%$ C.L.
Comparing this result with the sensitivity of JUNO, in Eq. \eqref{eq:Limits_CC}, it can be concluded that JUNO can provide constraints about an order of magnitude more stringent than HK.

Similarly, we have investigated the HK sensitivity for the axion-photon and axion-nucleon couplings. In the small mass limit ($m_a\lesssim 10$~keV), solar axions would be detected via inverse Primakoff absorption on oxygen. Utilizing Eq. \eqref{Counts_PC}, replacing $N_{C}$ with the number of oxygen $N_{O}\simeq 1.25 \times 10^{34}$,
the HK sensitivity  at $90\,\%$ C.L.reads
\begin{equation}
| g_{3aN} \times g_{a\gamma} | \lesssim 1.18 \times 10^{-11}\,{\rm GeV}^{-1}\qquad {\rm for}\,\,m_a\lesssim 10~{\rm keV}\;.
\label{eq:Limits_PC}
\end{equation} 
Even in this case, we find that JUNO can explore couplings about an order of magnitude smaller than HK.  
Indeed, though its exposure is lower than HK, JUNO has better sensitivity due to its excellent energy resolution [Cf. Eqs.~\eqref{eq:eres} and \eqref{eq:ResHK}].

Let us conclude by mentioning that the future Deep Underground Neutrino Experiment (DUNE) detector, which is currently under construction and expected to start taking data in the early-2030's, is less suitable to detect 5.49 MeV axions due to its high energy threshold ($E_{\rm th}\gtrsim 5$~MeV)~\cite{DUNE:2020ypp} and a higher background due to natural radioactivity in the surrounding rock and due to the charged current interaction of solar neutrinos on argon \cite{Capozzi:2018dat,Zhu:2018rwc}.

\section{Conclusions}
\label{sec:Conclusion}
In this work we have investigated the sensitivity of the neutrino detector JUNO to probe 5.49 MeV solar axions produced in the $p(d,^3{\rm He})a$ reaction. 
The possible detection through Compton conversion would allow JUNO to probe the combination $| g_{3aN} \times g_{ae} | \gtrsim 6.33 \times 10^{-14}$ at $90\,\%$ C.L. for $m_a\lesssim 1~{\rm MeV}$. 
For larger masses, axions can decay into electron-positron pairs and the JUNO sensitivity reaches $|g_{3aN}\times g_{ae}| \sim 10^{-8}$. 
On the other hand, due to the inverse Primakoff process JUNO will explore the combination $ | g_{3aN} \times g_{a\gamma} | \gtrsim 6.5 \times 10^{-12}\,{\rm GeV}^{-1}$ for $m_a\lesssim 10~{\rm keV}$, while for larger masses the axion decay into photons leads to the sensitivity   $| g_{3aN} \times g_{a\gamma} |\times m_a^2 \lesssim 3.3 \times 10^{-12}\,\,{\rm eV}$. 

Due to its large exposure time and the excellent energy resolution, JUNO will be able to set the strongest experimental limits on the combinations $|g_{3aN}\times g_{ae}|$ and $|g_{3aN}\times g_{a\gamma}|$, improving by more than one order of magnitude the Borexino bounds, and it has the best sensitivity among the current and proposed neutrino experiments, such as Hyper-Kamiokande. 

 Our study has shown an example of the physics potential of large underground neutrino detectors in probing axions. Other studies could include the evaluation of the Super-Kamiokande sensitivity to detect muonphilic axions produced from charged-meson decays in air showers~\cite{Cheung:2022umw} and the search for cosmogenic relativistic axions with future neutrino detectors, such as HK and JUNO itself~\cite{Cui:2022owf}. This connection deserves further investigations to complement the standard experimental techniques to study axions.

\section*{Acknowledgements}
We warmly thank Eligio Lisi for helpful discussions and comments on the manuscript, as well as Davide Franco for clarifying some aspects of the Borexino analysis in the first stage of this work. 
The work of G.L. and 
A.M. is partially supported by the Italian Istituto Nazionale di Fisica Nucleare (INFN) through the ``Theoretical Astroparticle Physics'' project
and by the research grant number 2017W4HA7S
``NAT-NET: Neutrino and Astroparticle Theory Network'' under the program
PRIN 2017 funded by the Italian Ministero dell'Universit\`a e della
Ricerca (MUR).  N.N. is supported by the Istituto Nazionale di Fisica Nucleare (INFN) through the “Theoretical Astroparticle Physics” (TAsP) project.  

\appendix 
\section{Supernova bound}
\label{app:supernova}
In this Appendix, we present a short discussion of the SN bound on the axion-nucleon coupling $g_{3aN}$. 
In general, SN 1987A provides one of the most stringent bounds on the axion-nucleon couplings. 
Axions with mass up to $\lesssim $ 100 MeV, as the ones considered in this work, can be thermally produced in a SN and, if their couplings are sufficiently weak, they stream out without being reabsorbed.
This, in turn, could dramatically alter the predictions for the observed neutrino signal from SN 1987A~\cite{Raffelt:1987yt,Turner:1987by,Mayle:1987as}.

Here, we consider the most updated SN bound~\cite{Carenza:2019pxu}, which assumes the nucleon-nucleon bremsstrahlung production of axions, $NN \to NN a$.\footnote{Refs.~\cite{Carenza:2020cis,Fischer:2021jfm} showed that the production through scattering on negative pions would  be more efficient production mechanism. However, no explicit bound was presented in this case. We will ignore this possibility here.} 
We further assume that the axion-nucleon coupling is small enough to allow them to escape from the SN (free streaming regime). 
The exact free streaming threshold is quite difficult to calculate but we can assume $ g_{3aN}\sim10^{-6}$~\cite{Carenza:2019pxu}.

The bound in Ref.~\cite{Carenza:2019pxu} applies to a specific combination of the axion coupling to neutrons ($g_{an}$) and protons ($g_{ap}$) 
$$f(g_{an},g_{ap})<8.26\times 10^{-19}\,,$$
where 
$$f(g_{an},g_{ap})=g_{an}^2+0.61\,g_{ap}^2+0.53\,g_{an}g_{ap}\,.$$

To translate this bound into a constraint on $g_{3aN}$, we define  $x=g_{ap}/g_{an}$ and express $g_{an}$ in terms of the effective coupling $g_{3aN}=(g_{ap}-g_{an})/2 \Rightarrow g_{an}=2g_{3aN}/(1-x)$. So we get
$$f=\frac{4\,g_{3aN}^2}{(1-x)^2}(1+0.53\,x+0.61x^2) \, .$$
The function $f$ has a minimum for $x=-2.53/1.75$, corresponding to $f\simeq g_{3aN}^2$. 
Thus, we find 
\begin{eqnarray}
g_{3aN}<9.1\times 10^{-10} \,.
\end{eqnarray}
This value represents the lower limit of the SN cooling bound on $g_{3aN}$ shown as the pink region in Fig.~\ref{fig:gAegA3N} and Fig.~\ref{fig:gagmev}.

\bibliographystyle{utphys}
\bibliography{SolarAxionV3}

\end{document}